\documentclass[aps,prl,twocolumn,superscriptaddress,nofootinbib,showpacs]{revtex4}
\newcommand{\apsi}
{\affiliation{Paul Scherrer Institute, CH-5232 Villigen PSI, Switzerland}}
\newcommand{\aucb}
{\affiliation{University of California, Berkeley, and LBNL, Berkeley, CA 94720, USA}}
\newcommand{\apnpi}
{\affiliation{Petersburg Nuclear Physics Institute, Gatchina 188350, Russia}}
\newcommand{\auiuc}
{\affiliation{University of Illinois at Urbana-Champaign, Urbana, IL 61801, USA}}
\newcommand{\aucl}
{\affiliation{Universit{\'e} Catholique de Louvain, B-1348, Louvain-la-Neuve, Belgium}}
\newcommand{\auk}{\affiliation{University of Kentucky, Lexington, KY 40506, USA}}
\newcommand{\aub}{\affiliation{Boston University, Boston, MA 02215, USA}}
\newcommand{\is}{s$^{-1}\;$}
\newcommand{\isn}{s$^{-1}$}

\usepackage{hyperref, amsmath, graphicx, verbatim}

\begin{document}

\title{Measurement of the Rate of Muon Capture in Hydrogen Gas and \\
Determination of the Proton's Pseudoscalar Coupling ${\textsl g}^{}_P$}
\apnpi
\aucb
\auiuc
\aucl
\apsi
\auk
\author{V.A.~Andreev}
\apnpi
\author{T.I.~Banks}
\aucb
\author{T.A.~Case}
\aucb
\author{D.B.~Chitwood}
\auiuc
\author{S.M.~Clayton}
\auiuc
\author{K.M.~Crowe}
\aucb
\author{J.~Deutsch}
\aucl
\author{J.~Egger}
\apsi
\author{S.J.~Freedman}
\aucb
\author{V.A.~Ganzha}
\apnpi
\author{T.~Gorringe}
\auk
\author{F.E.~Gray}
\aucb
\author{D.W.~Hertzog}
\auiuc
\author{M.~Hildebrandt}
\apsi
\author{P.~Kammel}
\auiuc
\author{B.~Kiburg}
\auiuc
\author{S.~Knaack}
\auiuc
\author{P.A.~Kravtsov}
\apnpi
\author{A.G.~Krivshich}
\apnpi
\author{B.~Lauss}
\aucb
\author{K.L.~Lynch}
\aub
\author{E.M.~Maev}
\apnpi
\author{O.E.~Maev}
\apnpi
\author{F.~Mulhauser}
\auiuc
\apsi
\author{C.S.~{\"O}zben}
\auiuc
\author{C.~Petitjean}
\apsi
\author{G.E.~Petrov}
\apnpi
\author{R.~Prieels}
\aucl
\author{G.N.~Schapkin}
\apnpi
\author{G.G.~Semenchuk}
\apnpi
\author{M.A.~Soroka}
\apnpi
\author{V.~Tishchenko}
\auk
\author{A.A.~Vasilyev}
\apnpi
\author{A.A.~Vorobyov}
\apnpi
\author{M.E.~Vznuzdaev} 
\apnpi
\author{P.~Winter}
\auiuc
\collaboration{MuCap Collaboration}
\date{\today}

\begin{abstract}
The rate of nuclear muon capture by the proton has been measured using 
a new experimental technique based on a time projection chamber 
operating in ultra-clean, deuterium-depleted hydrogen gas 
at 
1~MPa pressure.  
The capture rate was obtained from the difference between the measured $\mu^-$
 disappearance rate in hydrogen
and the world average for the $\mu^+$ decay rate.
The target's low gas density of 1\% compared to liquid hydrogen 
is key to avoiding uncertainties that arise from the formation of muonic 
molecules.  The capture rate from the hyperfine singlet ground state of 
the $\mu p$ atom is measured to be 
\mbox{$\Lambda^{}_S=725.0 \pm 17.4$~\isn}, 
from which the induced pseudoscalar coupling of the nucleon, 
\mbox{${\textsl g}^{}_P(q^2=-0.88\,m_\mu^2)=7.3\pm1.1$}, is extracted.  
This result is consistent with theoretical predictions for ${\textsl g}^{}_P$ 
that are based on the approximate chiral symmetry of QCD\@. 
\end{abstract}

\maketitle

We report the first result of the MuCap experiment for the rate 
$\Lambda^{}_S$ of the semileptonic weak process of ordinary 
muon capture~(OMC) by the proton,
\begin{equation}
  \mu^- + p \rightarrow n + \nu_\mu ~.
  \label{eq:mup}
\end{equation}
This fundamental process, like neutron beta decay, involves the 
vector and axial-vector form factors ${\textsl g}^{}_V(q^2)$ and 
${\textsl g}^{}_A(q^2)$, which characterize the microscopic QCD 
structure of the nucleon in electroweak charged-current interactions.  
Due to its larger momentum transfer $q_0^2=-0.88\,m_\mu^2$, 
reaction~(\ref{eq:mup}) is also sensitive to the weak magnetic and 
pseudoscalar induced form factors, ${\textsl g}^{}_M(q^2)$ and 
${\textsl g}^{}_P(q^2)$.  Form factors ${\textsl g}^{}_V(q^2_0)$, 
${\textsl g}^{}_M(q^2_0)$ and ${\textsl g}^{}_A(q^2_0)$ are 
accurately determined by experimental data and Standard Model 
symmetries and contribute an uncertainty of only 0.46\% to 
$\Lambda^{}_S$~\cite{gA-clarification}.  
Process~(\ref{eq:mup}) provides the most direct probe of 
${\textsl g}^{}_P \equiv {\textsl g}^{}_P(q^2_0)$, the pseudoscalar 
coupling of the nucleon's axial current, which is by far the least 
well known of these form factors.

The form factor ${\textsl g}^{}_P(q^2)$ arises mainly 
from the coupling of the weak leptonic current to the nucleon via 
an intermediate pion, which generates a pole term that dominates 
at $q^2_0$.  Early theoretical expressions for ${\textsl g}^{}_P$ 
were derived using current algebra techniques; now ${\textsl g}^{}_P$ 
can be systematically calculated within heavy baryon chiral 
perturbation theory~(HBChPT) up to two-loop order~\cite{Kaiser:2003dr}.  
The precise QCD result 
${\textsl g}^{}_P=8.26\pm0.23$~\cite{Bernard:1994wn,Bernard:2001rs} 
follows from the basic concepts of explicit and spontaneous chiral 
symmetry breaking, and thus its experimental confirmation is an 
important test of QCD 
symmetries~[\citealp{Bernard:2001rs}--\citealp{Govaerts:2000ps}].

\begin{figure}[t]
  \begin{center}
  \includegraphics[scale=0.4, angle=90]{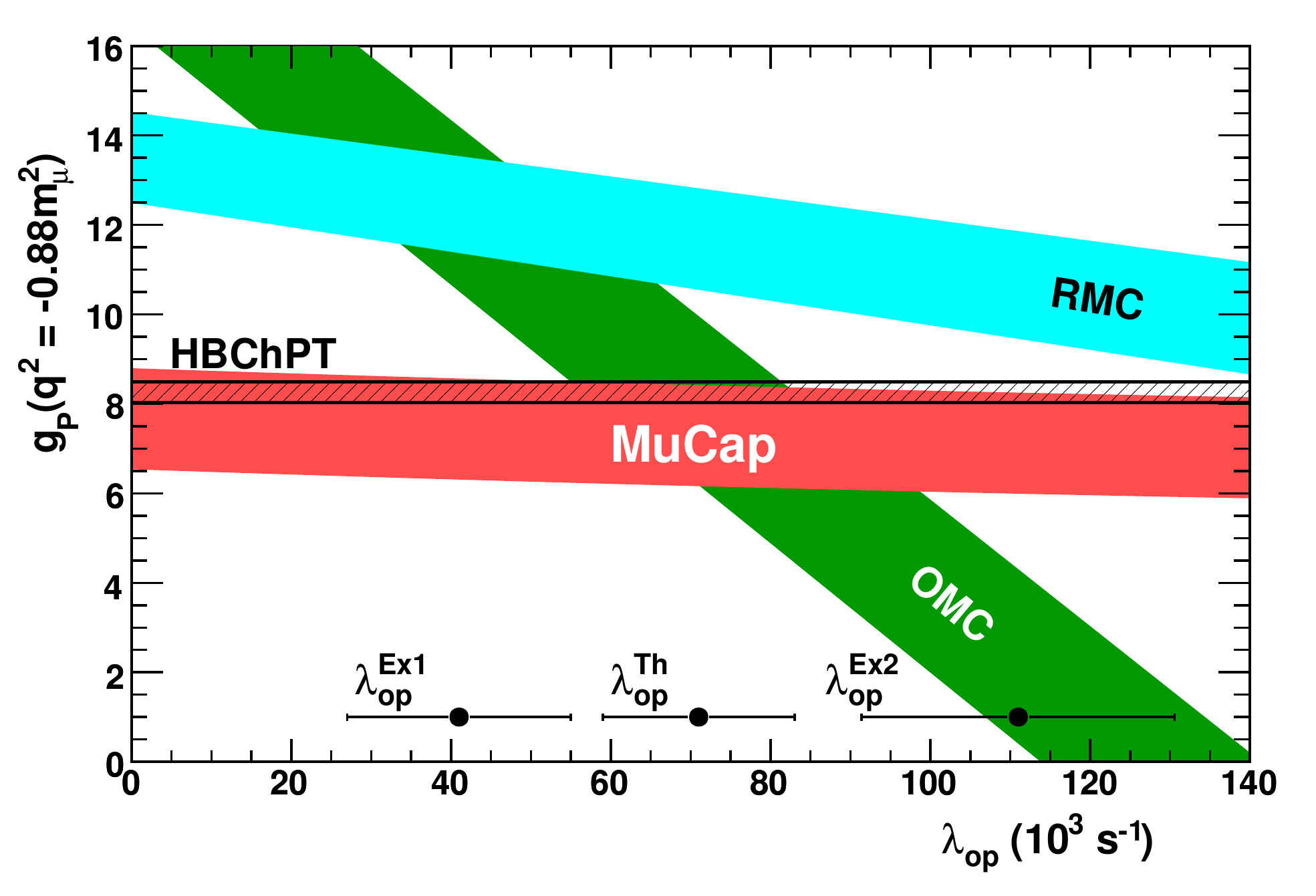}
  \vspace{-5mm}
  \caption{Experimental and theoretical determinations of 
  ${\textsl g}^{}_P$, presented vs.\ the ortho--para transition 
  rate $\lambda_{\rm op}$ in the $p\mu p$ molecule.  The most 
  precise previous OMC experiment~\cite{Bardin:1980mi} and the RMC
  experiment~\cite{Wright:1998gi} both depend significantly on the 
  value of $\lambda_{\rm op}$, which itself is poorly known due to 
  mutually inconsistent experimental 
  ($\lambda^{\rm Ex1}_{\rm op}$~\cite{Bardin:1981cq},
  $\lambda^{\rm Ex2}_{\rm op}$~\cite{Clark:2005as}) and theoretical 
  ($\lambda^{\rm Th}_{\rm op}$~\cite{Bakalov:1980fm}) results.  
  In contrast, the MuCap result for ${\textsl g}^{}_P$ is nearly 
  independent of molecular effects.}
  \label{g_P.fig}
  \end{center}
\vspace{-7mm}
\end{figure}

Experimental OMC efforts span a period of more than forty years, 
and more recently radiative muon capture~(RMC) by the proton was
measured for the first time~\cite{Wright:1998gi}.  
However, as shown in Fig.~\ref{g_P.fig}, the situation prior to the 
present experiment was 
inconclusive, as the results 
lacked sufficient precision due to ambiguities in the interpretation 
as well as technical challenges.

The problems of interpretation can be appreciated by considering the 
chain of reactions possible for negative muons after stopping in a
hydrogen target of density $\phi$ relative to liquid hydrogen
(LH$^{}_2$)~\cite{Gorringe:2002xx}.  Stopped muons immediately form 
ground state $\mu p$ atoms whose hyperfine states are populated in a 
statistical manner.  The upper triplet spin state is rapidly 
depopulated in collisions with H$^{}_2$ molecules, and for densities 
$\phi \geq 0.01$ all muons reach the $\mu p$ singlet state well 
before 100~ns.  From there, muons can either decay with 
a rate close to 
$\lambda_\mu^+ \equiv 1/\tau_\mu^+ \approx 0.455\times 10^6$~\isn, 
or be captured via reaction~(\ref{eq:mup}) at the predicted rate 
$\Lambda^{}_S\approx710$~\isn.  Complications arise at higher 
densities, however, as $\mu p$ atoms increasingly collide with target 
H$^{}_2$ molecules to form $p\mu p$ molecules.  The $p\mu p$ 
molecules are initially created in the ortho state at the 
density-dependent rate $\phi\lambda_{\rm of}$, where 
$\lambda_{\rm of}\approx 2.3\times 10^6$~\isn, and then de-excite to 
the para state at rate $\lambda_{\rm op}$.  The nuclear capture rates 
from the ortho and para states,
\mbox{$\Lambda_{\rm om} \approx 506$ \isn} and 
\mbox{$\Lambda_{\rm pm} \approx 208$ \is \cite{Bernard:2001rs}},
are quite different from each other and from $\Lambda^{}_S$, 
so knowledge of the relative populations of the $\mu p$ and 
$p\mu p$ states under any particular set of experimental 
conditions is crucial for a correct determination of 
${\textsl g}^{}_P$.  Alas, $\lambda_{\rm op}$ is poorly 
known~[\citealp{Bardin:1981cq}--\citealp{Bakalov:1980fm}]. 
This prevents a clear interpretation of the most precise OMC 
experiment~\cite{Bardin:1980mi}, which was performed in LH$^{}_2$ 
where muon capture occurs predominantly in $p\mu p$ molecules.  
The RMC process is less sensitive to $\lambda_{\rm op}$, but the 
large molecular uncertainties 
make it difficult 
to draw firm conclusions from the RMC experiment~\cite{Wright:1998gi}, 
whose results initially suggested a nearly 50\% higher value for
${\textsl g}^{}_P$ than predicted.

Direct measurement of $\Lambda^{}_S$ is technically difficult 
because process~(\ref{eq:mup}) is rare (branching ratio $= 0.16$\%) 
and leads to an all-neutral final state.  Moreover, target impurities 
and muon stops in detector walls must be scrupulously avoided, as 
negative muons preferentially and irreversibly transfer from $\mu p$ 
to heavier elements, and the nuclear muon capture rate increases 
roughly proportional to $Z^4$. The two previous muon 
capture experiments using low-density gas targets and neutron 
detectors obtained a precision in $\Lambda^{}_S$ of 
9\%~\cite{AlberigiQuaranta:1969ip} and 
13\%~\cite{Bystritsky:1974}, respectively.

The MuCap experiment employs novel techniques to minimize or avoid 
many of the problems described above.  The measurement is performed 
using hydrogen at density $\phi=(1.12\pm0.01)\times10^{-2}$, where 
$p \mu p$ formation is slow and 96\% of all captures proceed from the 
$\mu p$ singlet state.  
The significant background 
from muon stops in wall materials, inherent when using a low-density 
target, is eliminated by reconstructing the muon stopping point  
in an active target 
consisting of a hydrogen time projection chamber~(TPC).  
The capture rate is determined using the lifetime 
technique~\cite{Bardin:1980mi}, that is, from the difference between 
the measured disappearance rate 
$\lambda_\mu^- \approx \lambda_\mu^+ + \Lambda^{}_S$ 
of negative muons in hydrogen and 
the $\mu^+$ decay rate~$\lambda_\mu^+$, where it is assumed that free $\mu^-$ and $\mu^+$ 
decay with identical rates according to the CPT theorem.

\begin{figure}[t]
  \begin{center}
  \includegraphics[scale=0.27]{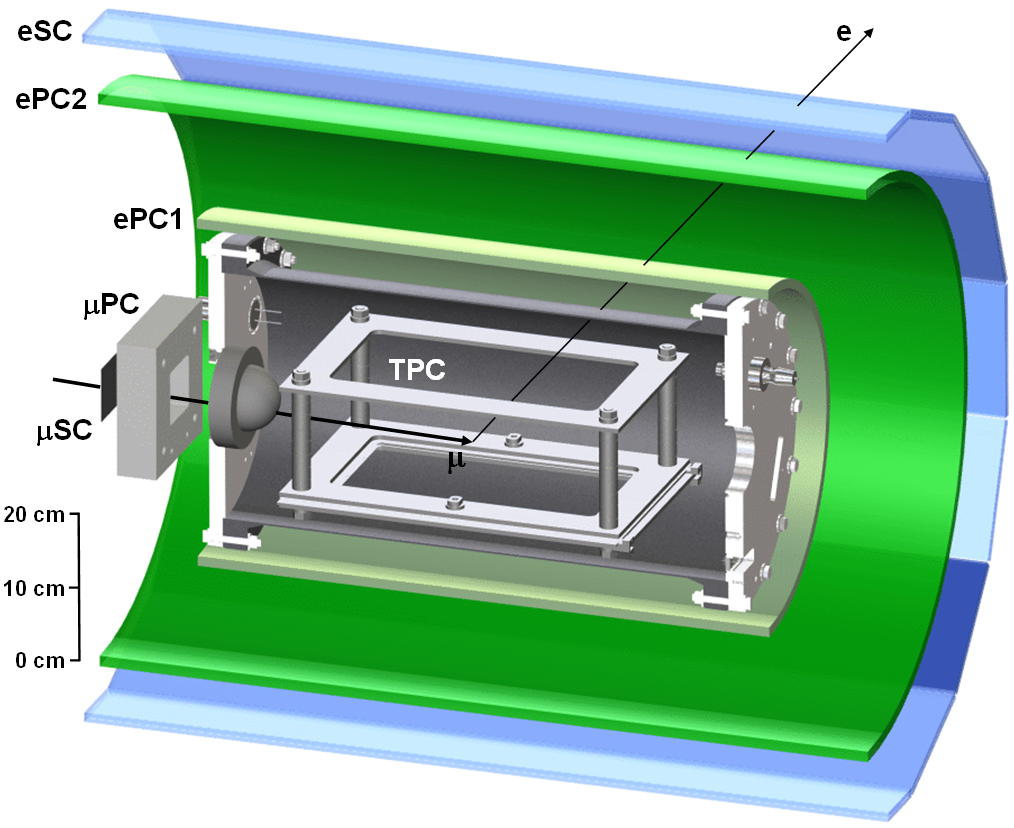}
  \caption{Simplified cross-sectional diagram of the MuCap detector.
   The detector components are described in the text.}
  \label{setup.fig}
  \end{center}
\vspace{-10mm}
\end{figure}
The experiment was conducted in the $\pi$E3 beamline at the Paul Scherrer 
Institute, using a $\approx20$~kHz DC muon beam tuned to a central momentum 
of $32.6$~MeV/$c$.  As illustrated in Fig.~\ref{setup.fig}, incident muons 
first traverse a plastic scintillator~($\mu$SC) and a multiwire proportional 
chamber~($\mu$PC), and then pass through a 0.5-mm-thick hemispherical 
beryllium window to enter an aluminum pressure vessel filled with 
ultra-pure, deuterium-depleted hydrogen gas at a pressure of 1.00~MPa and 
at ambient room temperature.  In the center of the vessel is the TPC 
(sensitive volume $15\times12\times28$~cm$^3$), which tracks incoming 
muon trajectories and thus enables the selection of muons that stop in the 
gas at least 15~mm away from chamber materials.  
Approximately 65\% of the muons passing through the $\mu$SC stop within 
this fiducial volume.  The ionization electrons produced by incoming muons 
drift downwards at velocity 5.5~mm/$\mu$s in an applied field of 2~kV/cm, 
towards a multiwire proportional chamber containing perpendicular anode
and cathode wires.  The anode plane consists of wires with 25~$\mu$m diameter and 4~mm
spacing, and a high voltage of 5.0~kV across the 3.5~mm half-gaps 
achieves a moderate gain of 60 in hydrogen.  Digital signals from 
three-level discriminators are recorded, with the energy thresholds 
adjusted to trigger on (i)~fast muons, (ii)~the Bragg peaks near the muon 
stopping points, and (iii)~the larger energies that may be deposited by 
recoiling nuclei following muon capture by gas impurities.

The TPC is surrounded by two cylindrical wire chambers~(ePC1, ePC2), each 
containing anodes and inner/outer cathode strips, and by a hodoscope 
barrel~(eSC) consisting of 16 segments with two layers of 5-mm-thick 
plastic scintillator.
This tracking system 
detects outgoing decay electrons with $3\pi$ solid angle acceptance.  All 
data are recorded in a triggerless, quasi-continuous mode to avoid deadtime 
distortions to the lifetime spectra.  Custom-built time-to-digital 
converters~(TDCs) digitize hit times for the TPC and the electron wire 
chambers.  The muon and electron times $t_\mu$ and $t_e$ are established by 
the  $\mu$SC and eSC detectors, and recorded in separate 
CAEN V767 TDC modules.

All TPC materials were carefully selected for high vacuum operation.
Prior to the run, the TPC system was heated to 115$^\circ$C under vacuum 
for several weeks to remove impurities.  
The system was filled 
with deuterium-depleted hydrogen through a palladium filter to remove
impurities.  During data taking, the gas was continuously 
circulated via an adsorption cryopump system and cleaned by 
cooled Zeolite filters~\cite{chups}, which achieved an equilibrium 
concentration (by number) of $c^{}_Z<5 \times 10^{-8}$, as monitored by 
direct TPC detection of recoil nuclei from muon capture by impurities.  
Gas chromatography measurements established that the atomic concentration 
of nitrogen was below $10^{-8}$, and the post-run installation of a 
humidity sensor with $10^{-9}$ sensitivity into the gas circuit indicated 
that the primary contaminant was H$^{}_2$O outgassing from within the 
pressure vessel.

\begin{figure}[t]
  \begin{center}
  \includegraphics[width=\columnwidth]{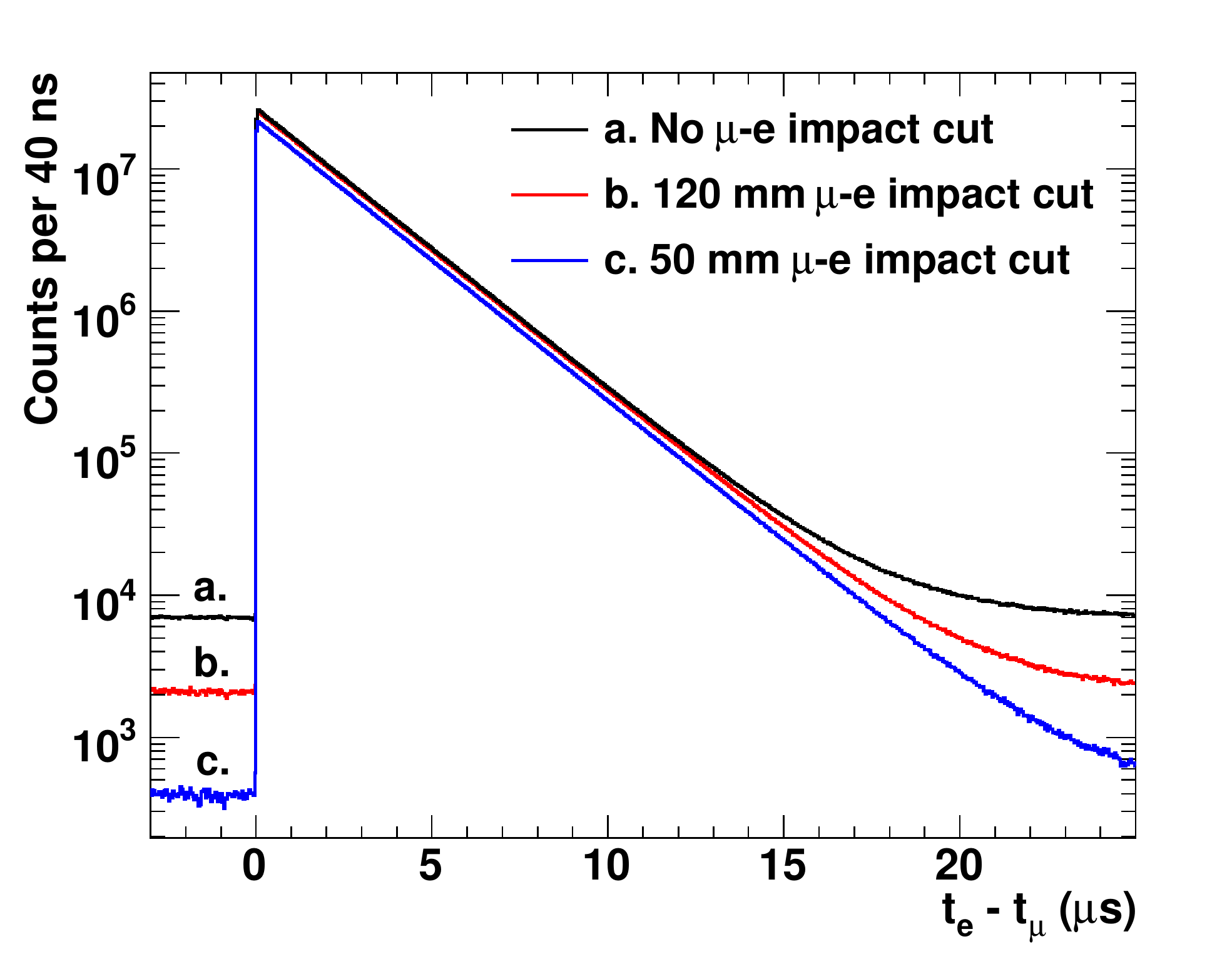}
  \caption{Lifetime spectra of negative muons.  The signal-to-background 
   ratio improves with tighter cuts on the $\mu$-e vertex.}
  \label{lifetime.fig}
  \end{center}
\vspace{-8mm}
\end{figure}

The isotopic purity of the hydrogen is critical.  Muons preferentially
transfer from $\mu p$ to $\mu d$ at the rate $\phi c_d \lambda_{pd}$, 
where $c_d$ is the deuterium concentration and 
$\lambda_{pd}\approx1.4\times10^{10}$~\isn.  Whereas $\mu p$ diffusion 
is on the order of mm, $\mu d$ atoms can diffuse cm-scale distances due 
to a Ramsauer-Townsend minimum in the $\mu d + p$ elastic scattering 
cross-section.  As a result, $\mu d$ atoms can drift sufficiently far 
away from the muon's original stopping point that the decay event will 
be rejected by the $\mu$-e vertex reconstruction cut in a time-dependent 
manner.  In addition, $\mu d$ atoms can drift into surrounding materials 
and be captured there.
Our target gas was produced 
via electrolysis of deuterium-depleted water, 
and accelerator mass spectrometry~(AMS)
measurements~\cite{AMS} determined that $c_d=(1.44\pm0.13)\times 10^{-6}$, 
roughly 100 times below deuterium's natural abundance. This result
was independently confirmed from our data, by analyzing the observed losses 
of muon decay events as a function of the imposed $\mu$-e vertex cut.


The time differences between muon arrivals and decay electron emissions, 
$\Delta t=t_{e}-t_{\mu}$, are histogrammed into lifetime spectra
(Fig.~\ref{lifetime.fig}). Only muons that are separated in time by 
$\pm$25~$\mu s$ from other muon arrivals are accepted.  While this 
condition cuts the usable statistics by $\approx 68$\%, it is essential
for avoiding systematic distortions to 
the background which can arise from ambiguities in resolving multiple 
muon tracks in the TPC, and it
dramatically improves the signal-to-background ratio. 
As shown in Fig.~\ref{lifetime.fig}, further background suppression can 
be achieved by performing a vertex cut on the impact parameter between 
each decay electron's trajectory and its parent muon's stopping point.  
In the final analysis we employ a loose impact parameter cut of 120~mm
as an optimal compromise between the competing demands for a good 
signal-to-background ratio and minimization of losses due to $\mu d$ 
diffusion out of the cut volume.

We fit the $\mu^-$ lifetime spectra with the simple exponential function
$f(t) = N w \lambda e^{-\lambda t} + B$, where the free parameters are 
the number of reconstructed decay events $N$, the disappearance rate 
$\lambda$, and the accidental background level $B$; $w$ is the fixed 
40~ns histogram bin width.  We studied 
an assortment of 
analysis conditions, including different time ranges (0.1--24~$\mu$s is 
typical), fiducial cuts, and detector combinations, and typically obtained 
 $\chi^2/$dof = 0.95--1.02 for 600 degrees of freedom. 

\begin{table}
\begin{tabular}{lr@{}lr@{}l}
\hline\hline
Source                  & \multicolumn{2}{c}{Correction (\isn)}	
                        & \multicolumn{2}{c}{Uncertainty (\isn)} \\ 
\hline
$Z>1$ impurities        & \hspace*{0.23in} $-$19&.2             
                        & \hspace*{0.37in} 5&.0 \\
$\mu d$ diffusion       & $-$10&.2             &  1&.6 \\
$\mu p$ diffusion       &  $-$2&.7             &  0&.5 \\
$\mu + p$ scattering    & & 
                                                &  3&   \\
$\mu$ pileup veto efficiency & & 
                                                &  3&   \\
Analysis methods        & & 
                                                &  5&   \\
\hline
Total                   &  $-$32&.1             &  8&.5 \\
\hline\hline
\end{tabular}
\caption{Systematic corrections and uncertainties applied
to the observed $\mu^-$ disappearance rate $\lambda$.}
\label{sys_table}
\vspace{-3mm}
\end{table}

In reality, the experimental $\mu^-$ lifetime spectrum is not a pure 
exponential, but 
has a more complicated shape due to 
contributions from $p\mu p$ molecules and hydrogen gas impurities.  
However, these effects are sufficiently small that their perturbations 
$\Delta\lambda$ to the exponential decay rate $\lambda$ are linear and 
can be corrected sequentially.  
The main corrections to $\lambda$ were derived directly from 
experimental data, with some additional information from external 
measurements and literature.  For residual $c^{}_Z$ below a few times
$10^{-6}$, the correction $\Delta\lambda^{}_Z$ scales with the observed 
impurity capture yield per muon, $Y^{}_Z$, as
$\Delta\lambda^{}_Z = Y^{}_Z \Bigl[
\alpha^{}_{\rm N} 
\bigl( \frac{\Delta\lambda_{\rm N}}{Y_{\rm N}} \bigr)^{\rm calib} + 
\alpha^{}_{\rm O} 
\bigl( \frac{\Delta\lambda_{\rm O}}
{Y_{\rm O}} \bigr)^{\rm calib} \Bigr]$.
The observed yield $Y^{}_Z \approx 11 \times 10^{-6}$,
received contributions from nitrogen and humidity in weights 
of approximately $\alpha^{}_{\rm N}=0.05$ and 
$\alpha^{}_{\rm O}=0.95$.
The factors $(\frac{\Delta\lambda}{Y})^{\rm calib}$ were empirically 
fixed by calibration runs involving N$^{}_2$ and O (in the form of 
H$^{}_2$O) concentrations 50--1000 times above their values in the 
clean fill.  We find $\Delta\lambda^{}_Z=-19.2\pm 5.0$~\isn, where 
the error is dominated by a conservative estimate of the 
$(\frac{\Delta\lambda}{Y})^{\rm calib}$ value for O,
determined during our 2006 running period.
The correction for deuterium-related diffusion effects, 
$\Delta\lambda_d=-10.2\pm1.6$~\isn, was obtained by a 
zero-extrapolation procedure using data from a run with a hydrogen 
filling of $c_d=(122 \pm 5) \times 10^{-6}$. 

The preceding corrections are summarized in Table~\ref{sys_table}.  
There we also present four additional sources of uncertainty,
including a conservative error of 5~\is that accounts for the spread 
in results observed for a variety of consistency studies, as 
performed by two independent analyses.
To prevent bias, the master clock (accurate to $10^{-8}$~\cite{MuLan}), 
was detuned by an offset, which was concealed 
until the data analysis was complete.

The final result for the $\mu^-$ disappearance rate in pure hydrogen,
based on $N=1.6\times 10^9$ fully tracked, pileup-protected decay 
events from our 2004 data set, is 
$\lambda_\mu^- = 
455\,851.4\,\pm\,12.5_{\rm stat}\,\pm\,8.5_{\rm syst}$~\isn.
As a consistency check, we also measured the  
$\mu^+$ decay rate from $N=0.5\times 10^9$ events to be
$\lambda_\mu^+ = 455\,164\,\pm\,28$~\isn, in agreement with the 
world average.

The observed $\mu^-$ disappearance rate can be written as\vspace{-2mm}
\begin{equation}
\lambda_\mu^- = \left( \lambda_\mu^+ + \Delta \lambda_{\mu p} \right)
                              + \Lambda^{}_S + \Delta \Lambda_{p\mu p} ~.
  \label{eq:lmu}
\end{equation}
Here $\Delta\lambda_{\mu p}=-12.3$~\is describes a small reduction 
in the muon decay rate in the bound $\mu p$ system~\cite{Uberall:1960}. 
The term $\Delta\Lambda_{p\mu p}=-23.5 \pm 4.3 \pm 3.9$~\is 
accounts for captures from $p\mu p$ molecules, and is calculated from 
the full $\mu^-$ kinetics in pure hydrogen.  Its 
error terms come from our estimates $\lambda_{\rm of}=(2.3 \pm 0.5)\times 10^6$~\is 
and $\lambda_{\rm op}= (6.9 \pm 4.3)\times 10^4$~\isn, respectively, which cover
most of the existing literature values. 
As muon capture
from the $\mu p$ singlet component dominates both in $\mu p$ atoms and $p\mu p$
molecules, $\Delta\Lambda_{p\mu p}$ implicitly 
depends on $\Lambda_S$, which leads to a 3.2\% loss in sensitivity 
when determining $\Lambda_S$ from Eq.~\eqref{eq:lmu}.
Using the new world average 
$\lambda_\mu^+ = 455\,162.2 \pm 4.4~{\rm s}^{-1}$~\cite{MuLan},
we determine the rate of muon capture by the proton to be
\begin{equation}
  \Lambda_S^{\rm MuCap} = 
     725.0 \,\pm\, 13.7_{\rm stat} \,\pm\, 10.7_{\rm syst} ~{\rm s}^{-1} ~. 
  \label{eq:Ls}
\end{equation}
To compare with theory we consider the two recent NNLO calculations 
of $\Lambda^{}_S$, 687.4~\isn~\cite{Bernard:2000et} and 
695~\isn~\cite{Ando:2000zw}, here averaged to 691.2~\isn.  
Adding the very recently calculated radiative correction 
$\Delta_R=19.4$~\isn~\cite{Czarnezki06} (increased from 
$\Delta_R=4.5$~\isn~\cite{Goldman:1973qu}) yields the value
$\Lambda_S^{\rm Th}=710.6$~\is and enables us to calculate
\begin{equation}
  {\textsl g}_P^{\rm \,MuCap} 
    = {\textsl g}_P^{\rm \,Th} + 
      \frac{\partial {\textsl g}^{}_P}{\partial \Lambda^{}_S} 
      \left( \Lambda_S^{\rm MuCap} - \Lambda_S^{\rm Th} \right)
    = 7.3 \pm 1.1 ~,
\label{eq:gp}
\end{equation}
where ${\textsl g}_P^{\rm \,Th}=8.26$~\cite{Bernard:2001rs}, 
$\frac{\partial {\textsl g}^{}_P}
{\partial \Lambda^{}_S}=-0.065$~s~\cite{Govaerts:2000ps},
and only the experimental uncertainty from Eq.~\eqref{eq:Ls} is 
propagated. 
The linear expansion in Eq.~\eqref{eq:gp} is valid because 
of the small difference $\Lambda_S^{\rm MuCap} - \Lambda_S^{\rm Th}$, 
but should be refined once further theoretical work clarifies the
present 1\% difference between calculations~\cite{Bernard:2000et} 
and~\cite{Ando:2000zw} and quantifies all sources of theoretical 
uncertainty at the sub-percent level.

The current information on ${\textsl g}^{}_P$
is summarized in Fig.~\ref{g_P.fig}; the constraints~\cite{Gorringe:2002xx} from the
OMC experiment~\cite{Bardin:1980mi} are updated
to reflect the larger $\Delta_R$~\cite{Czarnezki06}. 
The situation before MuCap was inconclusive and exhibited mutually inconsistent 
theoretical predictions and experimental determinations of both 
${\textsl g}^{}_P$ and $\lambda_{\rm op}$.  The low 
gas density in MuCap renders our result relatively insensitive to 
$\lambda_{\rm op}$ and thus avoids most model dependence, enabling 
us to report the first precise, unambiguous determination of 
${\textsl g}^{}_P$.  This experimental result agrees with present 
theory to within $1\sigma$ and does not support a dramatic deviation 
from the chiral prediction as the RMC result originally had implied.  
Additional data are being collected with the aim of a more than 
twofold reduction of statistical and systematic uncertainties.

We are grateful to the technical staff of the collaborating institutions 
for their vital contributions to the experiment, and notably to the PSI 
staff for delivering beam of excellent quality.  Thanks are due to 
A.~Adamczak,
L.~Bonnet,
R.M.~Carey,
P.T.~Debevec,
A.~Dijks\-man,
D.~Fahrni, 
A.A.~Fetisov, 
B.~Gartner,
J.~Govaerts,
F.J.~Hartmann, 
A.~Hofer, 
V.I.~Jatsoura, 
V.~Markushin,
L.~Meier,
D.~Michotte,
C.J.G.~Onderwater, 
J.~Paley, 
S.~Sadetsky, and 
P.A. Zolnierczuk
for their contributions to the development of the experiment, and to  
N.~Bondar and T.~Ferguson for providing CMS front-end electronics 
for the ePCs.  This work was supported in part by the U.S. 
National Science Foundation, the U.S. Department of Energy and CRDF, 
PSI, the Russian Academy of Sciences, and a grant of the President of 
the Russian Federation (NSH-3057.2006.2).  Essential computing 
resources for the analysis were provided by the National Center for 
Supercomputing Applications.


\end{document}